\documentclass[%
 reprint,
 amsmath,amssymb,
 aps,prl,showpacs
]{revtex4-1}

\usepackage{graphicx}
\usepackage{dcolumn}
\usepackage{bm}

\begin{document}


\title{Criticality and mechanical enhancement in composite fibre networks}

\author{Jan Maarten van Doorn}
\author{Luuk Lageschaar}%
\author{Joris Sprakel}
\author{Jasper van der Gucht}
 \email{jasper.vandergucht@wur.nl}
\affiliation{%
 Physical Chemistry and Soft Matter, Wageningen University, Stippeneng 4, 6708 WE, Wageningen, The Netherlands
}%

\date{\today}

\begin{abstract}
Many biological materials  consist of sparse networks of disordered fibres, embedded in a soft elastic  matrix. The interplay between rigid and soft elements in such composite networks leads to mechanical properties that can go far beyond the sum of those of the constituents. Here we present lattice-based simulations to unravel the microscopic origins of this mechanical synergy. We show that the competition between fibre stretching and bending and elastic deformations of the matrix gives rise to distinct mechanical regimes, with phase transitions between them that are characterized by critical behaviour and diverging strain fluctuations and with different mechanisms leading to mechanical enhancement.  
\end{abstract}
\maketitle
Many materials, ranging from textiles and paper to connective tissue and the cytoskeleton of living cells, have a microscopic structure that consists  of crosslinked  fibres. Theoretical progress in the last decades has led to a detailed understanding of the physics of such fibre networks\cite{Broedersz2014}. Because stiff fibres resist not only stretching, but also bending, the mechanical behaviour of fibre networks differs significantly from that of networks of flexible polymers. Different mechanical regimes can be observed: at high densities fibre networks deform affinely and the elasticity is governed by fibre stretching, while at lower densities  there is a crossover to a non-affine, bending-dominated regime\cite{Wilhelm2003,Head2003,Head2003b,Das2007,Broedersz2011}.\\
\indent Although experiments on model networks give support to the existence of different mechanical regimes\cite{Gardel2004,Lieleg2007,Sharma2016}, the current theories fall short in describing real biomaterials. An important reason for this is that natural materials are almost without exception composite materials that consist of mixtures of elements of different rigidity:  the cytoskeleton is a complex network of (partially bundled) actin filaments, intermediate filaments, and microtubules\cite{Bausch2006}; the extracellular matrix consists  of stiff collagen fibres in a matrix of more flexible polymers\cite{Wegst2004};  and also many synthetic high-performance materials are composites of soft and rigid fibres\cite{Munch2008,Wegst2015, Jang2015,Gong2003,Sun2012}. It is clear that the collective non-affine deformation modes that characterize the mechanics of sparse fibre networks must be hindered significantly by the presence of an elastic matrix\cite{Huisman2010,Bai2011,Zhang2013,Wada2009,Das2011}, but a fundamental understanding of how this interplay affects the mechanical properties of composites has remained elusive.\\ 
\indent Here we use numerical simulations to study the mechanics of disordered composite networks, consisting of crosslinked fibres embedded in a soft elastic matrix. Both the fibres and the polymers that constitute the background matrix are arranged on a 2D triangular lattice with lattice spacing $l_0$, as shown in Fig. 1. The effects of connectivity are explored by randomly removing segments of the fibre network with a probability $1-p$, so that the average connectivity equals $z=6p$. Sequences of contiguous colinear fibre segments are treated as elastic rods, characterized by a stretch modulus $\mu_1$ and a bending modulus $\kappa_1$. Since fibres in biomaterials are typically much softer with respect to bending than to stretching\cite{Broedersz2014}, we will only consider the case that $\kappa_1\ll\mu_1l_0^2$. Intersecting fibres are assumed to be crosslinked with permanent, but freely-hinged bonds. The background matrix is modelled as a homogeneous network of undiluted central force springs with stretch modulus $\mu_2$. The two networks are linked to each other at each vertex of the lattice. To investigate the mechanical response of the composite network, we calculate the linear shear modulus $G$  by applying a shear strain $\gamma$ to the boundaries and minimizing the total elastic energy by relaxing the internal degrees of freedom (see Supplementary Information).\\
\begin{figure}[hbt!]
\centering
\includegraphics[width=\linewidth]{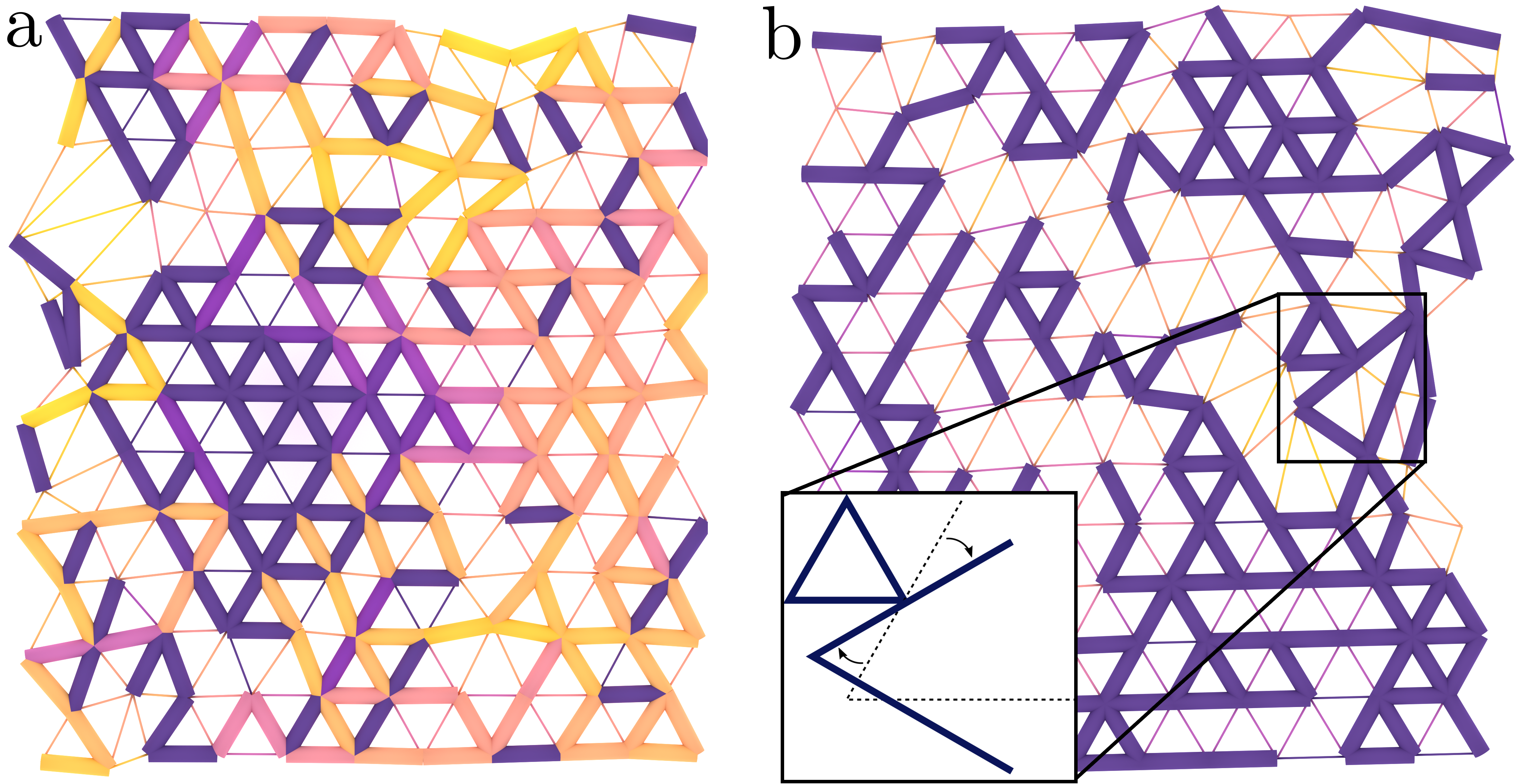}
\caption{Composite networks on a triangular lattice. A small section of a deformed network of fibres in a soft matrix, with $\kappa_1/(\mu_2l_0^2)=10^{-6}$ and $\mu_2/\mu_1=10^{-12}$ for (a) $p=0.65$ and (b) $p=0.45$. Thick segments represent fibre segments, color-coded for their bending energy (yellow: strongly bent, blue:  weakly bent),  and  thin segments represent the background matrix, color-coded for stretching energy (yellow: strongly stretched; blue: weakly stretched). Inset in b shows an example of a rigid rotation of a fibre cluster.}
\end{figure}
\newpage
\begin{figure}[hbt!]
\centering
\includegraphics[width=\linewidth]{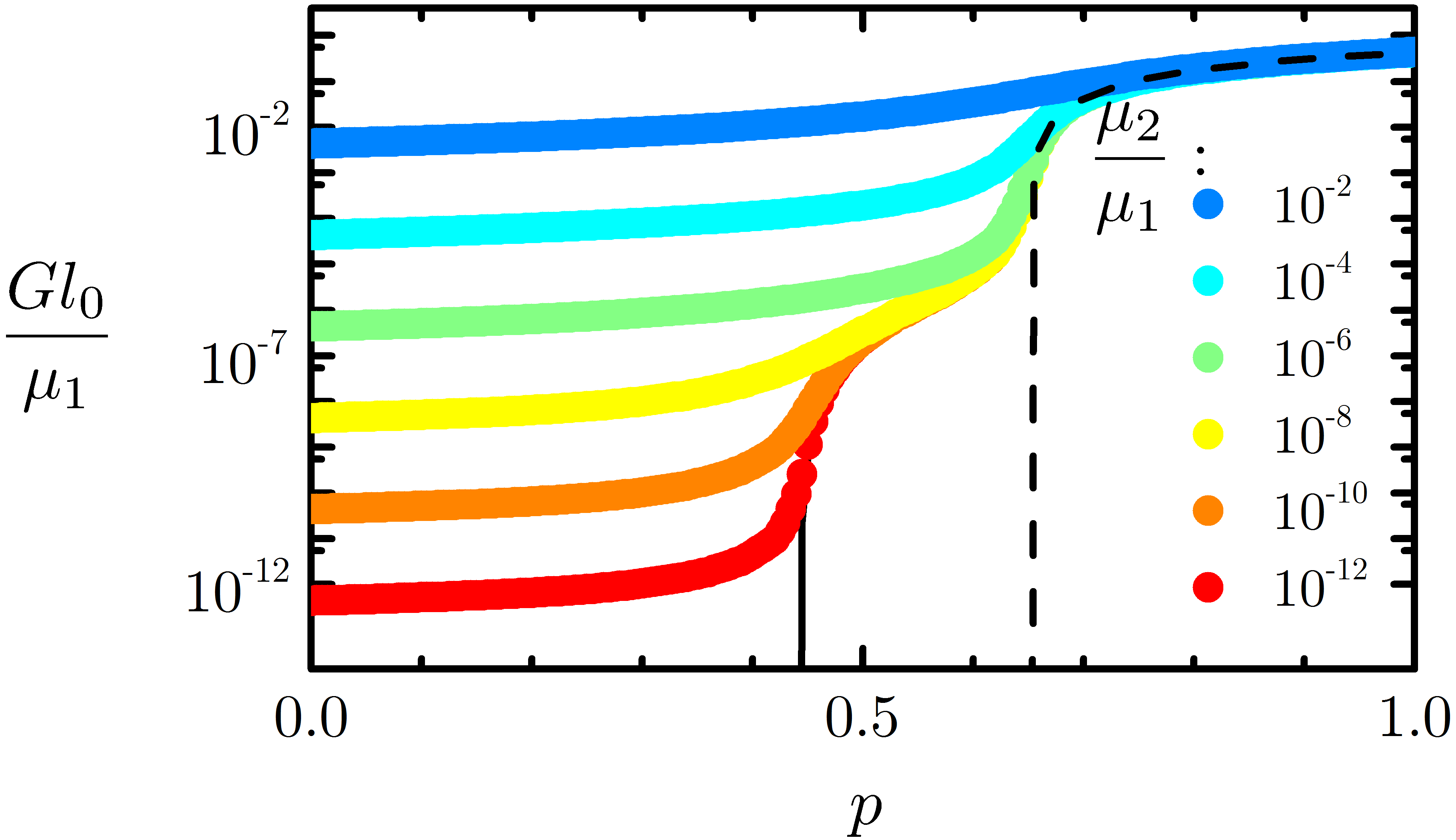}
\caption{Elasticity of composite networks. Shear modulus $G$ (in units $\mu_1/l_0$) as a function of the bond probability $p$ for $\kappa_1/\mu_1 l_0^2=10^{-6}$ and for a range of stiffnesses of the background matrix. The black line corresponds to $\mu_2=0$ and the dashed line to $\mu_2=0$ and $\kappa_1=0$. }
\end{figure}
\indent In Fig. 2, we show the shear modulus as a function of the connectivity $p$ for various values of the matrix stiffness $\mu_2$.  For $\mu_2=0$, $G$ vanishes when the connectivity is lower than a critical rigidity threshold. For fibres with no bending rigidity ($\kappa_1=0$, dashed line), this threshold is $p_\textrm{cf}\approx0.651$, as given by Maxwell's criterion for isostatic networks of central force springs\cite{Maxwell1864}. For non-zero $\kappa_1$, however, the rigidity threshold shifts discontinuously to a lower value, $p_\textrm{b}\approx0.442$, which is independent of $\kappa_1$ for $\kappa_1>0$ (black line, see also Supplementary Fig. 1)\cite{Broedersz2011,Mao2013}. 
In the presence of an elastic matrix with non-zero stretch modulus $\mu_2$, the network is mechanically stable for any value of $p$. However, features of the mechanical transitions at $p_\textrm{cf}$ and $p_\textrm{b}$ can still be seen, as the shear modulus decreases very steeply with decreasing $p$ around these points (Fig. 2). This suggests that both points mark a transition between distinct mechanical regimes in the composite network.  
\indent To investigate the nature of these different regimes, we examine both crossover regions in more detail.  
For low values of $\mu_2$, the mechanical response of the composite network is dominated by the fibre network for $p$ sufficiently above $p_\textrm{b}$. We therefore expect that the crossover region at $p_\textrm{cf}$ is similar to the one observed in single-component fibre networks. As shown previously\cite{Broedersz2011}, in such networks the central force threshold coincides with a transition from a stretching-dominated regime for $p>p_\textrm{cf}$ to a bending-dominated regime for $p<p_\textrm{cf}$.  The presence of an elastic matrix as embedding medium is  expected to affect this transition, because fibre bending is a non-affine deformation mode, which inevitably leads to additional strain in the medium. The elastic energy stored in the matrix due to the bending of an embedded fibre increases proportionally to the matrix stiffness $\mu_2$\cite{Brangwynne2006}. We therefore expect the resistance to bending to increase linearly with $\mu_2$. Indeed, we find that we can collapse our data by introducing an effective bending rigidity, which is the sum of the intrinsic bending rigidity and a matrix-induced bending resistance (see Supplementary Information):  
\begin{equation}\label{eq:keff}
\kappa_\textrm{eff}=\kappa_1+\mu_2 l_0^2
\end{equation}
This is shown in Fig. 3a, where we plot the scaling form
\begin{equation}\label{eq:scCF}
G=\frac{\mu_1}{l_0}|\Delta p_\textrm{cf}|^\beta \mathcal{G}^\textrm{cf}_{\pm}\left(\frac{\kappa_\textrm{eff}}{\mu_1l_0^2}|\Delta p_\textrm{cf}|^{-\alpha}\right)
\end{equation}
with $\Delta p_\textrm{cf}=p-p_\textrm{cf}$ and with scaling exponents $\alpha=3.0$ and $\beta=1.4$, in agreement with previous findings\cite{Broedersz2011}. The universal scaling function $\mathcal{G}^\textrm{cf}_{\pm}(x)$ consists of three branches that characterize three different mechanical regimes. For $x\ll1$, $\mathcal{G}^\textrm{cf}_{+}(x)\sim\textrm{const}$ and $\mathcal{G}^\textrm{cf}_{-}(x)\sim x$. This implies a stretching-dominated regime with $G\sim\mu_1 |\Delta p_\textrm{cf}|^{\beta}$  above the transition ($\Delta p_\textrm{cf}>0$), and a bending-dominated regime with $G\sim\kappa_\textrm{eff}|\Delta p_\textrm{cf}|^{\beta-\alpha}$ below the transition ($\Delta p_\textrm{cf}<0$). In the bending-dominated regime, the shear modulus is governed by the effective bending resistance of the fibres (equation \ref{eq:keff}): for very soft matrices ($\mu_2< \kappa_1l_0^{-2}$) the response is dominated by the intrinsic bending rigidity of the fibres, $G\sim\kappa_1$, while for stiffer matrices ($\mu_2> \kappa_1l_0^{-2}$) the shear modulus is determined by the induced bending rigidity due to the matrix: $G\sim\mu_2$.  Very close to the critical threshold, 
we find a cross-over regime with anomalous scaling\cite{Broedersz2011} $G\sim\kappa_\textrm{eff}^{\beta/\alpha}\mu_1^{1-\beta/\alpha}$ independent of $\Delta p_\textrm{cf}$, as observed from the critical branch in Fig. 3a. 

At $p=p_\textrm{b}$ there is a second transition, now from a bending-dominated regime to a matrix-dominated regime. Again, we can capture the different regimes around this transition by a scaling form 
\begin{equation}\label{eq:scB}
G=\frac{\kappa_1}{l_0^3}|\Delta p_\textrm{b}|^\delta \mathcal{G}^\textrm{b}_{\pm}\left(\frac{\mu_2 l_0^2}{\kappa_1}|\Delta p_\textrm{b}|^{-\gamma}\right)
\end{equation}
with $\Delta p_\textrm{b}=p-p_\textrm{b}$ and $\mathcal{G}^\textrm{b}_{\pm}(x)$ another universal scaling function. The data is found to collapse with critical exponents $\gamma=4.5$ and $\delta=3.0$. Again, we see three branches, corresponding to three different mechanical regimes. Above the transition for $x\ll 1$ we find $\mathcal{G}^\textrm{b}_{+}(x)\sim\textrm{const}$ and $G\sim\kappa_1|\Delta p_\textrm{b}|^{\delta}$, which corresponds to the rigidity percolation scaling of a bending-dominated network\cite{Broedersz2011}. Below  the transition, for $x\ll 1$ we find $\mathcal{G}^\textrm{b}_{-}(x)\sim x$ and $G\sim\mu_2|\Delta p_\textrm{b}|^{\delta-\gamma}$. In this regime the fibre network is below its rigidity threshold, and the composite network consists of an elastic matrix with embedded, non-percolating fibre clusters. Indeed, the scaling that we find is very similar to the one found for a central force network with rigid inclusions\cite{Sahimi1993,Garboczi1986}.  Very close to the transition we again find an anomalous scaling regime in which the modulus becomes independent of $\Delta p_\textrm{b}$ and is governed by both bending and matrix contributions, with $G\sim\kappa_1^{1-\delta/\gamma}\mu_2^{\delta/\gamma}$. 
The different mechanical regimes that we find for our composite network are summarized in the phase diagram in Fig. 3c, which clearly highlights the rich behaviour of composite networks.

\begin{figure}[hbt!]
\centering
\includegraphics[width=\linewidth]{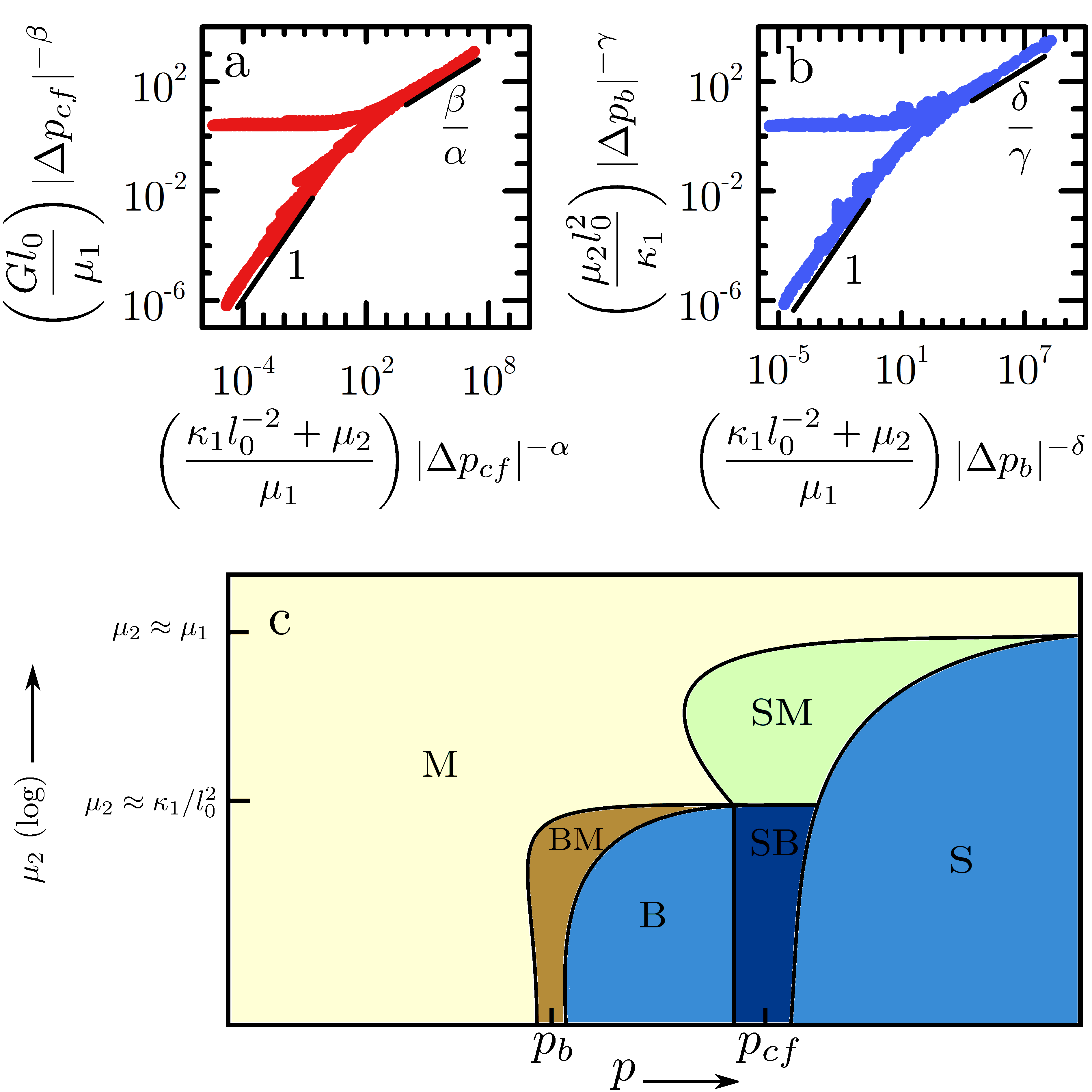}
\caption{Mechanical regimes in composite networks. Scaling analysis of the shear modulus in the vicinity of (a) the central force isostatic point $p_\textrm{cf}$  and (b) the rigidity threshold $p_\textrm{b}$, for a wide variety of values of $\kappa_1$ and $\mu_2$. Values of the critical exponents: $\alpha=3.0$, $\beta=1.4$, $\gamma=4.5$, $\delta=3.0$. (c) Mechanical phase diagram of composite networks: S: stretching-dominated ($G\sim\mu_1$), B: bending-dominated ($G\sim\kappa_1$), M: matrix-dominatd ($G\sim\mu_2$), SB: stretch-bend coupled ($G\sim\mu_1^{1-x}\kappa_1^x$), SM: stretch-matrix coupled ($G\sim\mu_1^{1-x}\mu_2^x$), BM: bend-matrix coupled ($G\sim\kappa_1^{1-y}\mu_2^y$).} 
\end{figure}
\begin{figure}[hbt!]
\centering
\includegraphics[width=\linewidth]{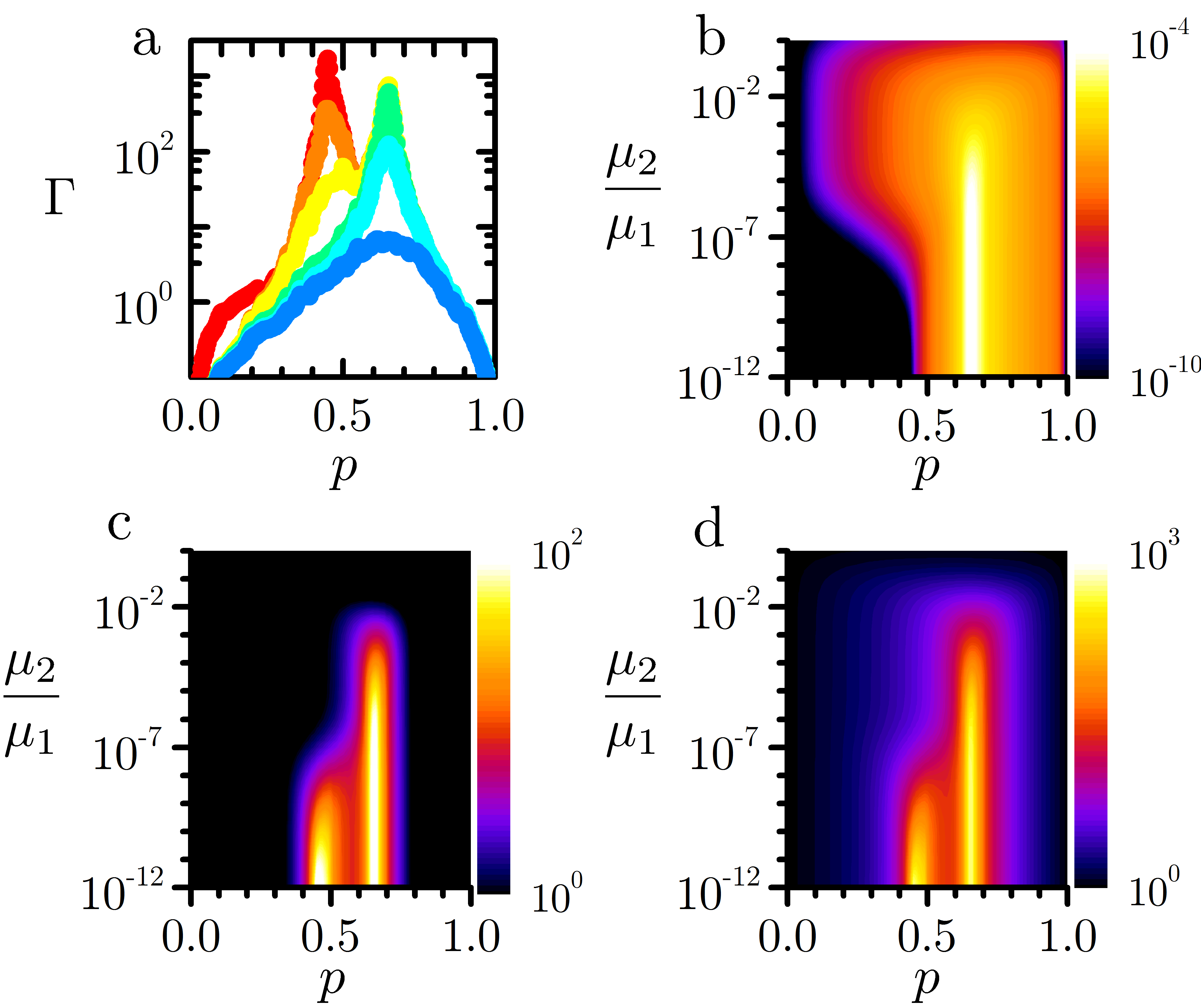} 
\caption{Non-affine deformations in composite networks. (a) Non-affinity as a function of connectivity $p$ for several values of $\mu_2$ (same colour coding as in Fig. 2).  (b) Bending energy per unit area and unit strain, $E_\textrm{b}/A\gamma^2$, as a function of $p$ and $\mu_2$. (c) Rigid body rotations: mean-squared rotation angle of the end-to-end vector of fibres, averaged over all fibres in the network $\langle\Delta\phi^2\rangle$, compared to that for the affinely deformed network as a function of $p$ and $\mu_2$. (d) Relative deformation energy of the background matrix, compared to the affinely deformed network, $E_2/E_2^\textrm{(aff)}$, as  a function of $p$ and $\mu_2$. The bending rigidity $\kappa_1=10^{-6}\mu_2l_0^2$ in all cases.} 
\end{figure}
\begin{figure*}[hbt!]
\centering
\includegraphics[width=0.8\paperwidth]{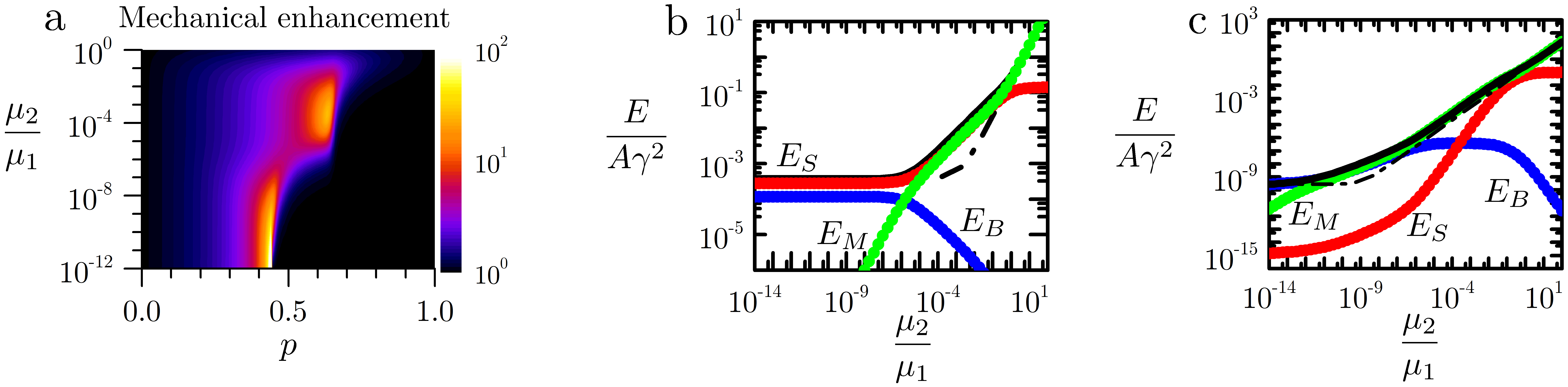} 
\caption{Mechanical enhancement in composites. (a) Enhancement of the shear modulus with respect to the summed moduli of the individual networks, $G/(G_1+G_2)$ as a function of $p$ and $\mu_2$ for $\kappa_1=10^{-6}\mu_2l_0^2$. (b,c) Different energy contributions  to the shear modulus ($E_\textrm{b}$ (blue): fibre bending; $E_\textrm{s}$ (red): fibre stretching; $E_\textrm{m}$ (green): matrix deformation) as a function of $\mu_2$ for (b) $p=0.65$ and (c) $p=0.45$. The black line gives the total elastic energy and the dashed line  the sum of the energies of the separate networks, so that the difference between the solid and the dashed line represents the mechanical enhancement. } 
\end{figure*}
It is well-established that the mechanics of weakly-connected disordered networks are governed by non-affine deformation modes\cite{Broedersz2014,Wilhelm2003,Head2003,Head2003b,Das2007,Broedersz2011}. This raises the question whether the different mechanical regimes that we observe originate from a transition between different non-affine modes. We examine the the non-affine fluctuations by calculating the mean-square deviation from a uniform, affine strain field\cite{DiDonna2005}:
\begin{equation}\label{eq:nonaffinity}
\Gamma=\frac{1}{\gamma^2 l_0^2}\left\langle\left(\mathbf{u}-\mathbf{u}^\textrm{(aff)}\right)^2\right\rangle
\end{equation}
Here $\mathbf{u}$ and $\mathbf{u}^\textrm{(aff)}$  are the actual displacement and the affine displacement of a node, respectively. We find a strong, cusp-like increase of the non-affine fluctuations in the vicinity of both $p_\textrm{cf}$ and $p_\textrm{b}$, highlighting the critical state of the fibre network at these points (Fig. 4a). From Fig. 1 it is clear, however, that the nature of the the non-affine modes is very different in these two regimes. For $p\approx p_\textrm{cf}$, the deformation field is characterized by large and heterogeneous bending fluctuations (Fig. 1a and 4b). This is in agreement with earlier work\cite{Wilhelm2003,Head2003,Head2003b,Das2007,Broedersz2011}, where the central force threshold was shown to mark a transition from an affine, stretching-dominated regime for $p>p_\textrm{cf}$ to a non-affine, bending-dominated regime for $p<p_\textrm{cf}$. By contrast, the increase in $\Gamma$ at $p\approx p_\textrm{b}$ is not associated with bending fluctuations (Fig. 1b), but can be ascribed to rigid body motions of fibres or fibre clusters (inset Fig. 1b and Fig. 4c) that become more and more prominent as the connectivity of the network decreases. At the rigidity threshold $p_\textrm{b}$, the fibre network becomes floppy and all the strain can be accommodated by such rigid body motions without elastic energy cost in the fibre network\cite{Broedersz2011,Mao2013,Heussinger2006}. 
However, while the non-affine modes are soft modes for the fibre network, they lead to additional  deformations in the background matrix, so that the elastic energy of the matrix is strongly increased in regions where the non-affine fluctuations are large (Fig. 4d). This means that the final deformation field in a composite network is a compromise between energy stored in the fibre network (which can be reduced by non-affine modes) and energy stored in the matrix (which is enhanced by non-affine deformations). As the matrix becomes stiffer, the non-affine fluctuations are increasingly suppressed (Fig. 4). The scaling of the non-affine fluctuations with $\mu_2$ and $\kappa_1$  are discussed in the Supplementary information (Supplementary Fig. 2).\\
\indent The main reason for the interest in composite materials is that the interplay between the different components can lead to highly synergistic properties, such as enhanced strength and rigidity\cite{Munch2008,Wegst2015, Jang2015,Gong2003,Sun2012}. We therefore consider the enhancement of the modulus of the composite network in comparison with the sum of the moduli of the individual networks (Fig. 5a). The highest enhancement, with a modulus that exceeds those of the individual networks by  up to a factor $10^2$, is observed in the two cross-over regions labelled SM and BM in Fig. 3c. We can understand the origin of the enhancement in these regimes,  by considering the different contributions to the modulus. At $p\approx p_\textrm{cf}$ (Fig. 5b), the modulus is dominated by bending contributions for small $\mu_2$. These  bending modes are suppressed by the elastic matrix when $\mu_2$ increases (Fig. 4b), leading to a more affine deformation. However, this goes at the cost of increased fibre stretching, and this increase in stretching energy stiffens the network. 
As discussed above, at $p\approx p_\textrm{b}$, the deformation of the fibre network is characterized by floppy modes, in which large clusters of fibres undergo rigid body motions without being strained. As the matrix becomes stiffer, these rigid body motions are suppressed at the cost of increased fibre bending (Fig. 4b,c).  Thus, while the enhancement around $p_\textrm{cf}$ is caused by the suppression of bending modes, the enhancement around $p_\textrm{b}$ is associated with an increase in fibre bending (Fig. 5c). 

We have revealed a very rich mechanical behaviour of composite networks. Small variations in composition can lead to large differences in mechanical response. This may be an important reason why composite structures are so abundant in biology, where adaptiveness is often crucial. Indeed, it has been argued that many biological networks have a connectivity in the vicinity of a critical regime\cite{Sharma2016}, where they are most susceptible to small changes. Our results show that these are also  the regions where mechanical synergy is to be expected. While our focus has been on linear elasticity, we expect that also the non-linear response of composite networks will differ greatly from that of single-component networks. Fibre networks are known to become stiffer as the strain increases\cite{Gardel2004} due a transition from bending to stretching-dominated elasticity\cite{Onck2005}. Recent experiments have shown that this strain stiffening can be suppressed completely when the fibres are embedded in a soft elastic matrix\cite{Rombouts2014}. Our results suggest that this may be the result of a suppression of bending modes already in the linear regime. Finally, the suppression of non-affine fluctuations by the background matrix leads to a more homogeneous stress distribution in the network. This should have large consequences for the nucleation and propagation of cracks in the material, and may thus contribute to the large increase in fracture strength found in double network hydrogels\cite{Gong2003,Sun2012}.\\

\section*{Acknowledgements}
This work is part of an Industrial Partnership Programme of the Foundation for Fundamental Research on Matter (FOM), which is part of the Netherlands Organisation for Scientific Research, and of the project SOFTBREAK funded by the European Research Council (ERC Consolidator Grant). 
\section*{References}

\end{document}